\documentstyle[preprint,aps,epsfig]{revtex}
\draft
\begin{document}
\preprint{\bf\it}
\title{Observable effects of symmetry energy in heavy-ion collisions 
at RIA energies}
\bigskip
\author{\bf Bao-An Li}
\address{Department of Chemistry and Physics\\
P.O. Box 419, Arkansas State University\\
State University, Arkansas 72467-0419, USA}
\maketitle

\begin{quote}
Within an isospin-dependent transport model for nuclear reactions induced by
neutron-rich nuclei, we perform a comparative study of $^{100}Sn+^{124}Sn$ 
and $^{100}Zn+^{124}Sn$ reactions at beam energies 
of 30 MeV/A and 400 MeV/A to identify optimal experimental conditions 
and observables for investigating the equation of state (EOS) of neutron-rich 
matter at RIA. Several observables known to be sensitive to the 
density-dependence of the symmetry energy are examined as a function of 
impact parameter for both reaction systems. In particular, the strength 
of isospin transport/diffusion is studied by using rapidity distributions of 
free nucleons and their isospin asymmetries. An approximate isospin equilibrium
is established even for peripheral collisions with a very soft symmetry 
energy at 30 MeV/A. At 400 MeV/A, however, a strong isospin translucency 
is observed even in most central collisions with either soft or stiff 
symmetry energies. Origins of these observations and their implications to 
determining the EOS of neutron-rich matter at the Rare Isotope Accelerator 
(RIA) are discussed. 
\end{quote}

\newpage
The isospin-dependence of the nuclear equation of state {\rm (EOS)} is one of
the most important but yet very poorly known properties of neutron-rich 
matter\cite{ibook}. Its determination in laboratory-controlled experiments
has profound implications to the study of many critical issues in 
astrophysics\cite{lat00,science}. Nuclear reactions induced by 
radioactive beams provide a unique opportunity to extract useful information
about the {\rm EOS} and novel properties of neutron-rich matter. 
In particular, at the Rare Isotope Accelerator {\rm (RIA)}, heavy-ions up to
400 MeV/A will be available. It just enables the study of neutron-rich
matter up to about 3 times normal nuclear matter density. A number of 
experimental observables have already been identified as sensitive probes 
of the EOS of neutron-rich matter
\cite{li97,fra1,fra2,xu00,tan01,bar02,betty,lizhang,li00,Greco03,li02,cor,cluster,tsang}. 
To gauge the sensitivity of some of these 
observables in reactions to be available at RIA, we carry out a comparative 
study of nuclear reactions nduced by $^{100}Sn (N/Z=1)$ and $^{100}Zn (N/Z=2.3)$ on 
$^{124}Sn (N/Z=1.48)$ target at 30 MeV/A and 400 MeV/A from central to 
peripheral collisions. Results of this study are useful for planning 
experiments and designing new detectors at RIA. Moreover, the role of isospin
degree of freedom in nuclear dynamics at RIA energies is an interesting 
subject in its own right. 

At present, nuclear many-body theories predict vastly different isospin
dependence of the nuclear {\rm EOS} depending on both the calculation 
techniques and the bare two-body and/or three-body interactions 
employed, see e.g., \cite{wir88,akm97,brown00,hor00}. Various theoretical 
studies (e.g., \cite{bom91,hub93}) have shown 
that the energy per nucleon $e(\rho,\delta)$ in nuclear matter 
of density $\rho$ and isospin asymmetry parameter $\delta$ defined as
$\delta\equiv(\rho_n-\rho_p)/(\rho_n+\rho_p)$
can be approximated very well by a parabolic function
$e(\rho,\delta)= e(\rho,0)+E_{sym}(\rho)\cdot\delta^2$.
In the above $e(\rho,0)$ is the {\rm EOS} of isospin 
symmetric nuclear matter 
and $E_{sym}(\rho)$ is the symmetry energy at density $\rho$. 
We shall use for isospin-symmetric nuclear matter a 
stiff {\rm EOS} with $K_0=380$ MeV which can reproduce the transverse 
flow data equally well as a momentum-dependent soft
{\rm EOS} with $K_0=210$ MeV\cite{pan93,zhang93}. 
The choice of $K_0$ does not affect our conclusions.
The {\rm EOS} of asymmetric matter should also be 
momentum-dependent since both the isoscalar and isovector potentials 
are momentum dependent\cite{das,li04}. 
For this comparative study, however, it is perfectly 
sufficient and much more efficient in calculations 
to use the momentum-independent {\rm EOS}. 
\begin{figure} 
\centering \epsfig{file=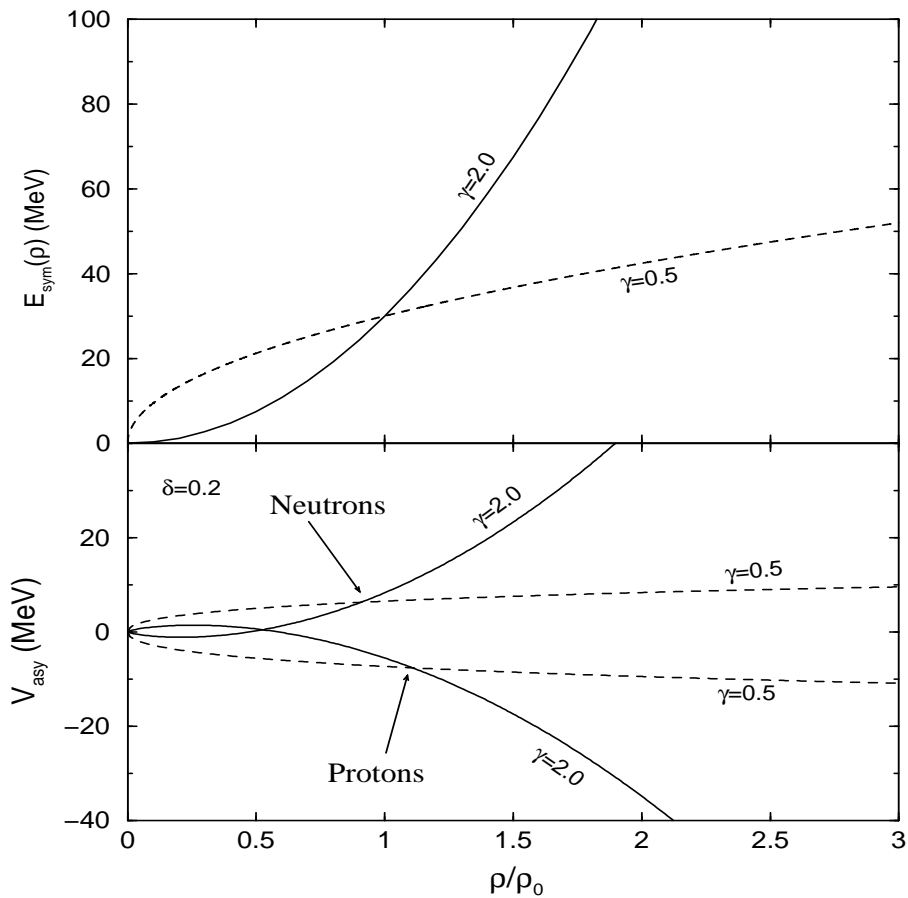,width=12cm,height=12cm,angle=-90}
\vspace{1 cm} 
\caption{Symmetry energy (upper window) and symmetry potentials 
as a function of density for an isospin asymmetry of $\delta=0.2$
and the $\gamma$ parameter of 0.5 and 2.0, respectively.} 
\label{fig1} 
\end{figure}     
The form of the symmetry energy as a function 
of density is rather strongly model 
dependent. We adopt here a parameterization used by Heiselberg 
and Hjorth-Jensen in their studies on neutron stars\cite{hei00}
$E_{sym}(\rho)=E_{sym}(\rho_0)\cdot u^{\gamma}$,
where $u\equiv \rho/\rho_0$ is the reduced density and $E_{sym}(\rho_0)$ is 
the symmetry energy at normal nuclear matter density $\rho_0$. 
By fitting the result of variational many-body calculations 
by Akmal et al\cite{akm97}, Heiselberg and Hjorth-Jensen found the values of 
$E_{sym}(\rho_0)=$32 MeV and $\gamma=0.6$. However, as shown by many other authors 
previously\cite{brown00,bom01} using other 
approaches, the extracted value of $\gamma$ varies widely, even its 
sign at supranormal densities is undetermined. In a recent 
analysis of isospin diffusion in heavy-ion 
collisions at 50 MeV/A at the NSCL/MSU, some experimental indications
of a $\gamma$ parameter as high as 2 was found\cite{tsang}. 
Therefore, in this work we study influence of the symmetry energy 
on experimental observables by varying the $\gamma$ parameter 
between 0.5 and 2. Assuming no momentum-dependence, 
the corresponding symmetry potentials can be derived directly from the 
density-dependent symmetry energy\cite{lizhang}. 
Shown in Fig.\ 1 are the symmetry energy (upper window) and symmetry
potentials (lower window) at $\delta$=0.2 as a function of density.  
It is seen that the soft ($\gamma=0.5$) symmetry energy leads to a larger (smaller) 
magnitude of the symmetry potential than the stiff one ($\gamma=2$) 
at densities below (above) about $\rho_0$. Since the magnitude of the
generally repulsive (attractive) symmetry potential for neutrons (protons) 
affects significantly the n/p ratio of emitted nucleons,
the stiffness of the symmetry energy is thus expected to affect 
the isospin asymmetry of emitted free nucleons differently at low 
and high energies. To investigate the density dependence of the symmetry
energy reactions with a broad range of beam energies are useful.

Our study is based on an isospin-dependent Boltzmann-Uehling-Uhlenbeck (IBUU) 
transport model (e.g., \cite{lizhang,li97,li98}). We select $^{100}Sn$ 
and $^{100}Zn$ on the opposite boundaries of stability of mass number 100 
as projectiles on a stable $^{124}Sn$ target. The initial neutron and proton density 
distributions of the projectile nuclei were calculated by using the Hartree-Fock-Bogoliubov
method and were provided to us by J. Dobaczewski\cite{jaeck}. With these choices the
$^{100}Sn+^{124}Sn$ reaction is symmetric (highly asymmetric) in protons (neutrons) in the
projectile and target, while the $^{100}Zn+^{124}Sn$ reaction is approximately 
symmetric (highly asymmetric) in
neutrons (protons). Comparing rapidity distributions of neutrons (protons), 
for instance, of these two reactions will allow us to investigate the isospin 
transport and possible isospin equilibration. 

Shown in Fig. 2 and Fig. 3 are the distributions of free neutrons 
(left panels) and protons (right panels) as a function of scaled rapidity 
at a beam energy of 30 MeV/A and 400 MeV/A, respectively, 
at an impact parameter of 1, 5 and 7 fm. Here we have 
selected free nucleons as those having local densities less than $\rho_0/8$ at freeze-out. 
\begin{figure}[htp] 
\centering \epsfig{file=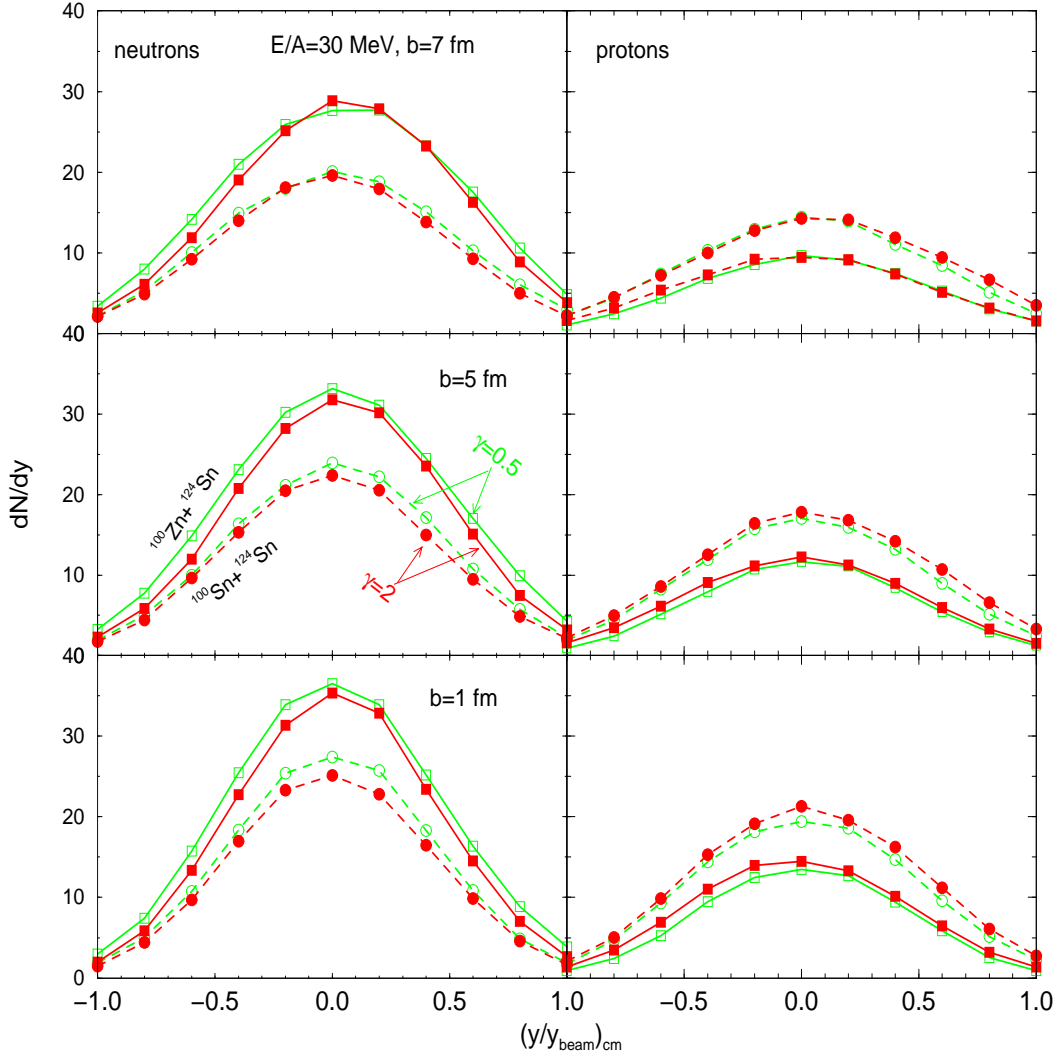,width=14cm,height=14cm,angle=-90} 
\vspace{1.0 cm} 
\caption{Rapidity distributions of neutrons (left) and protons (right) in the reaction
of $^{100}Sn+^{124}Sn$ and $^{100}Zn+^{124}Sn$ at a beam energy of 30 MeV/A.} 
\label{fig2} 
\end{figure}     

\begin{figure}[htp] 
\centering \epsfig{file=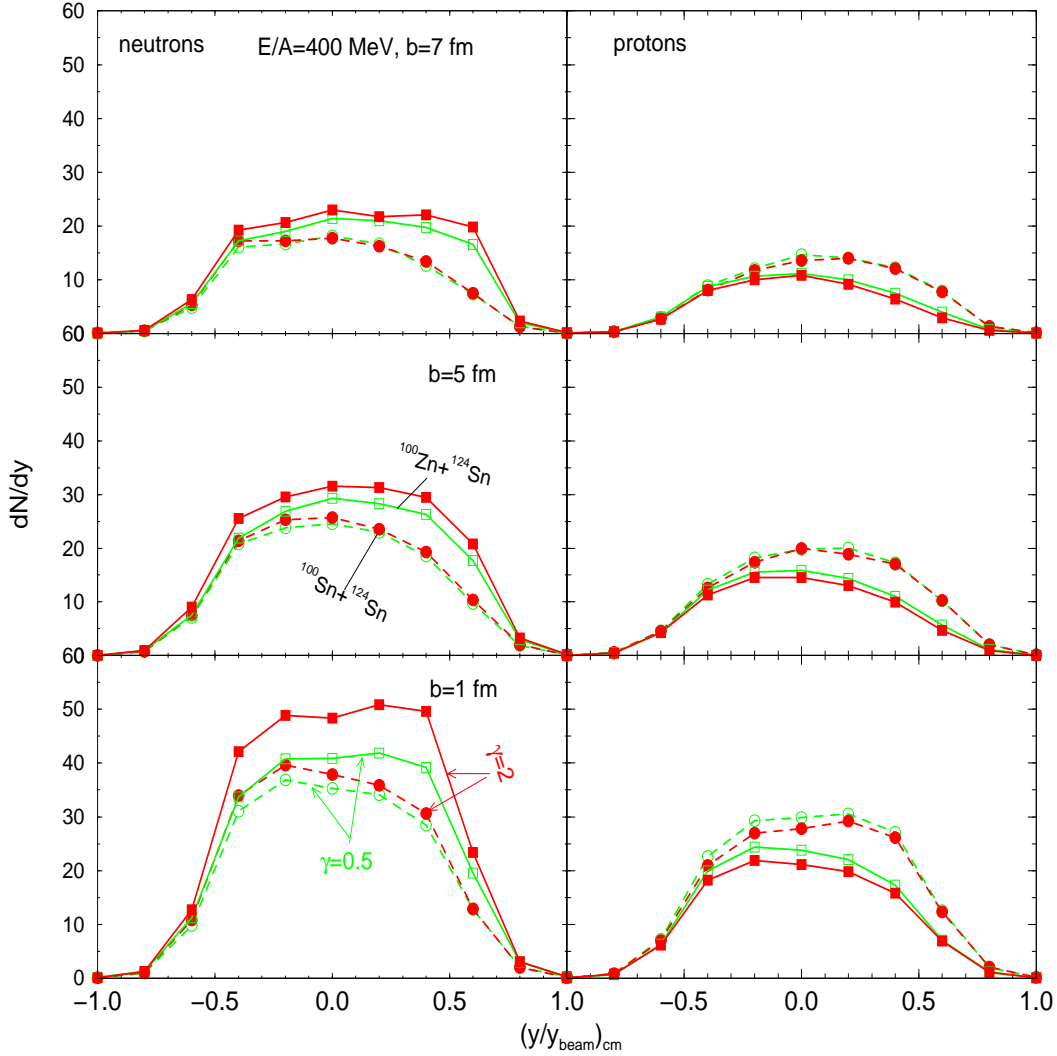,width=14cm,height=14cm,angle=-90}
\vspace{1.0 cm}  
\caption{Rapidity distributions of neutrons (left) and protons (right) in the reaction
of $^{100}Sn+^{124}Sn$ and $^{100}Zn+^{124}Sn$ at a beam energy of 400 MeV/A.} 
\label{fig3} 
\end{figure}     

First of all, it is seen that at 30 MeV/A the rapidity distributions 
of both neutrons and protons are about symmetric around $(y/y_{beam})_{cm}=0$, 
indicating a high degree of isospin and kinetic equilibrium.
At 400 MeV/A, however, distinct forward/backward asymmetries are seen in the rapidity
distributions of neutrons in $^{100}Sn+^{124}Sn$ and protons in  
$^{100}Zn+^{124}Sn$ even in the most central collisions, indicating a 
significant nuclear translucency and non-stopping. 
Secondly, the main effect of the generally repulsive (attractive)
symmetry potential for neutrons (protons) is to cause more (less) neutrons (protons) to
become free. This effect relies directly on the magnitude of the symmetry 
potential. At 30 MeV/A, since the symmetry potential is higher with $\gamma=0.5$ than with
$\gamma=2$ as shown in Fig. 1, there are thus more (less) free neutrons (protons) emitted
with $\gamma=0.5$ than with $\gamma=2$. While the opposite is true at 400 MeV/A since the 
magnitude of the symmetry potential reverses its order above $\rho_0$ with $\gamma=2$ 
leading to higher symmetry potentials now. 

\begin{figure}[htp] 
\centering \epsfig{file=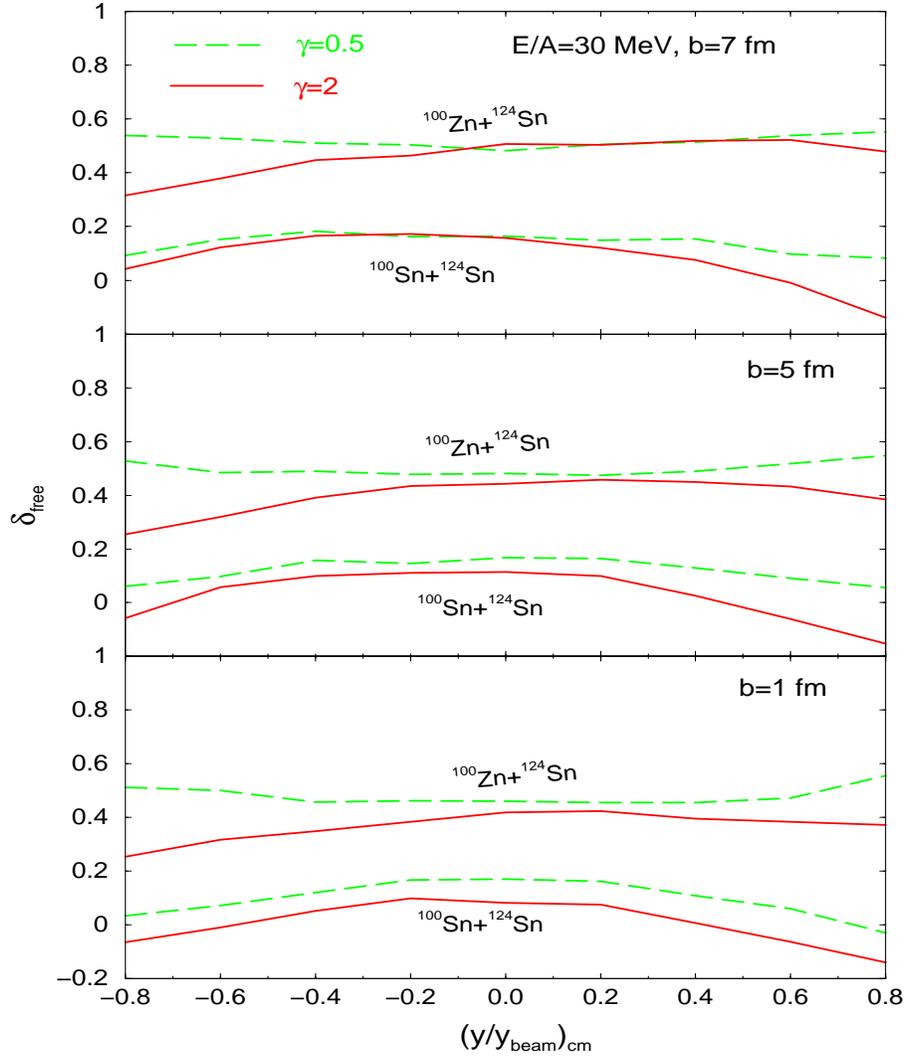,width=14cm,height=12cm,angle=-90} 
\vspace{1.0 cm} 
\caption{Rapidity distributions of isospin asymmetry of free nucleons in the reaction
of $^{100}Sn+^{124}Sn$ and $^{100}Zn+^{124}Sn$ at a beam energy of 30 MeV/A.} 
\label{fig4} 
\end{figure}     
\begin{figure}[htp] 
\centering \epsfig{file=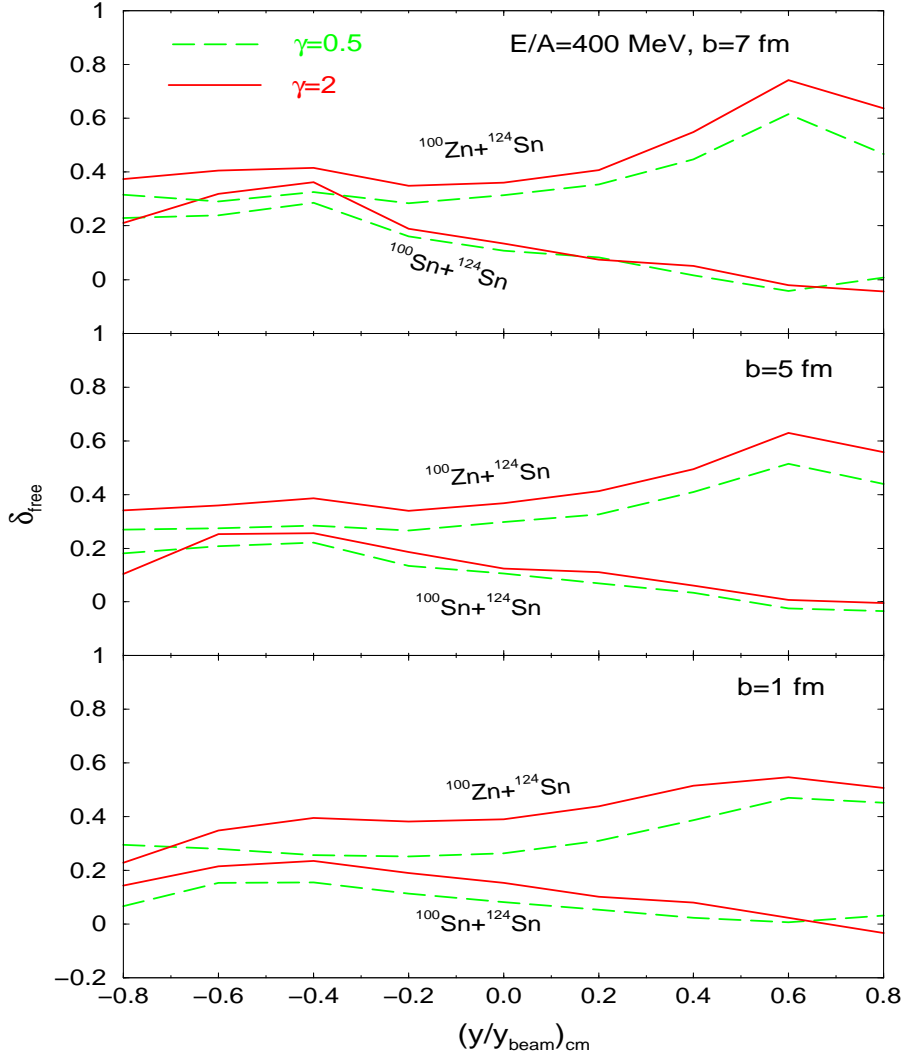,width=14cm,height=12cm,angle=-90}
\vspace{1.0 cm}  
\caption{Rapidity distributions of isospin asymmetry of free nucleons in the reaction
of $^{100}Sn+^{124}Sn$ and $^{100}Zn+^{124}Sn$ at a beam energy of 400 MeV/A.} 
\label{fig5} 
\end{figure}     
To be more quantitative about the degree of isospin equilibrium and to 
examine effects of the symmetry potential on isospin transport, we show in Fig. 4 
and Fig. 5 the isospin asymmetry of free nucleons as a function of rapidity.  
It is seen that a rather uniform isospin asymmetry is achieved over the whole rapidity range 
only with $\gamma$=0.5 at 30 MeV/A because of the large magnitude of the symmetry 
potential at densities relevant to the reaction. At 400 MeV/A, an appreciable backward-forward 
asymmetric isospin asymmetry exists for both reactions. With $\gamma=2$ the overall 
isospin asymmetry of free nucleons are significantly higher than that with $\gamma=0.5$. This is what 
one expects based on the density dependence of the symmetry potential shown in Fig. 1. 
Moreover, in contrast to the case at 30 MeV/A, the reaction time at 400 MeV/A is much 
too short compared to the time necessary for the system to reach the 
global isospin equilibrium\cite{lisherry,liko}.  

\begin{figure}[htp] 
\centering \epsfig{file=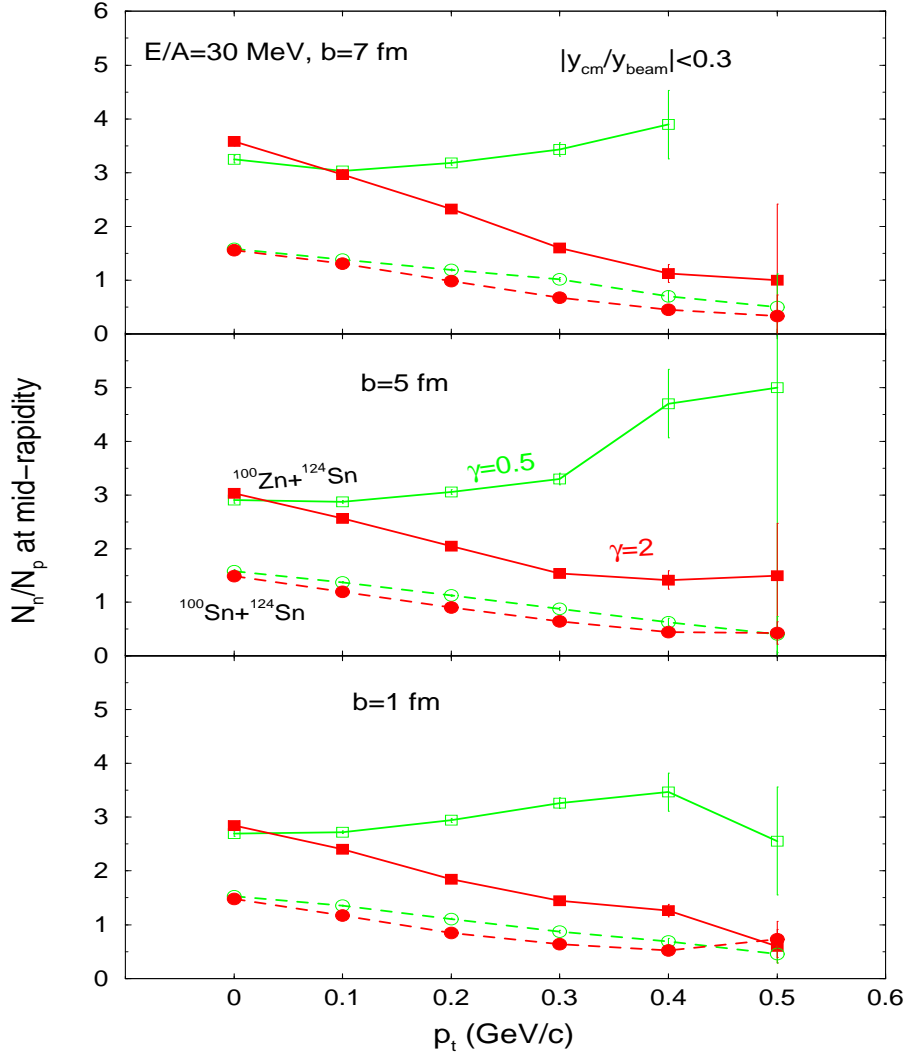,width=14cm,height=12cm,angle=-90}
\vspace{1.0 cm}  
\caption{Transverse momentum dependence of the neutron/proton ratio at mid-rapidity
for the reaction of $^{100}Sn+^{124}Sn$ and $^{100}Zn+^{124}Sn$ at a beam energy of 30 MeV/A.} 
\label{fig6} 
\end{figure}           
\begin{figure}[htp] 
\centering \epsfig{file=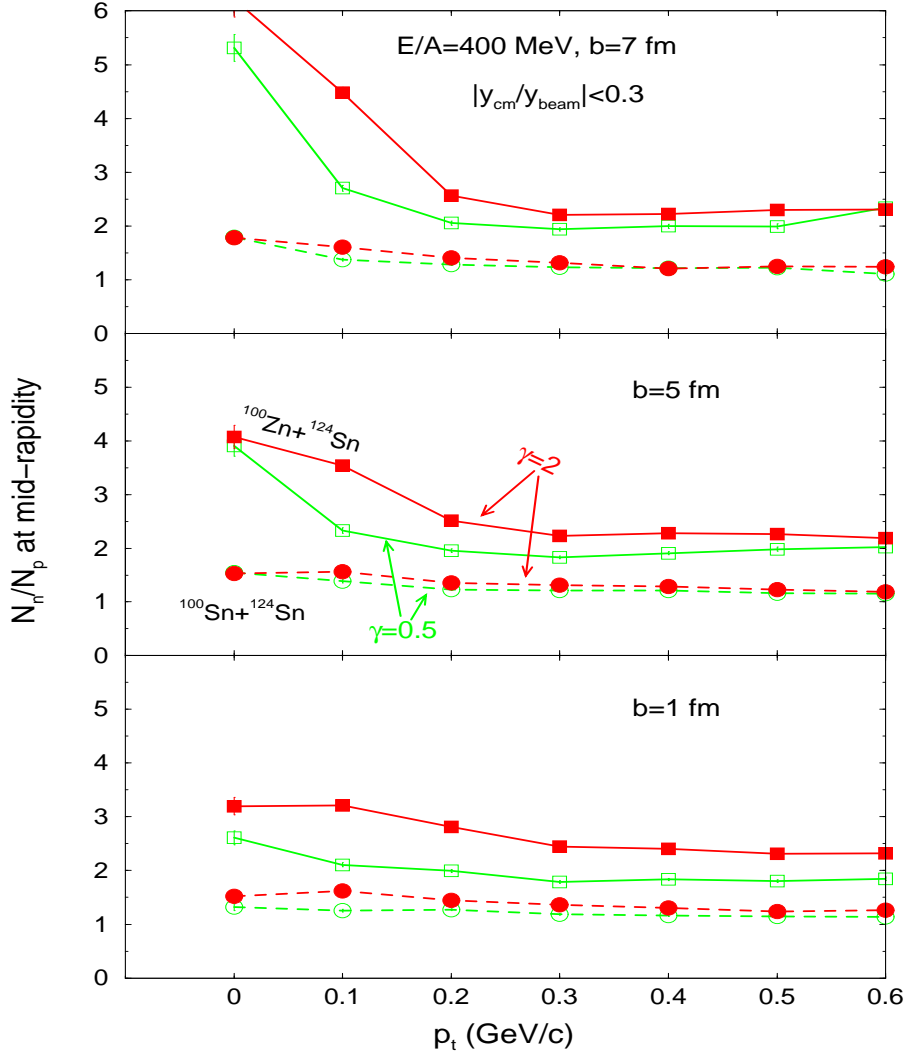,width=14cm,height=12cm,angle=-90}
\vspace{1.0 cm}  
\caption{Transverse momentum dependence of the neutron/proton ratio at mid-rapidity
for the reaction of $^{100}Sn+^{124}Sn$ and $^{100}Zn+^{124}Sn$ at a beam energy of 400 MeV/A.} 
\label{fig7} 
\end{figure}     

A good quantitative measure of the isospin transport is the neutron to proton 
ratio at mid-rapidity. Shown in Fig. 6 and Fig.7 are the n/p ratios at mid-rapidity versus 
the transverse momentum $p_t$ at 30 MeV/A and 400 MeV/A, respectively.
For the neutron-rich system $^{100}Zn+^{124}Sn$ at low transverse momenta 
at both beam energies, the n/p ratio is obviously increased compared to that 
of the reaction system. At both energies there is a general trend of an increasing n/p ratio at lower
transverse momenta, except the case of $\gamma=0.5$ at 30 MeV/A. This trend is mainly due to 
the Coulomb effect which shifts protons from low energies to higher energies. 
While the generally attractive symmetry potential for protons works against the 
Coulomb potential. For protons at low transverse momenta the Coulomb potential
dominates. While at higher transverse momenta with the parameter $\gamma=0.5$ the symmetry potential
dominates. With the soft symmetry energy at 30 MeV/A, the n/p ratio reaches the highest value 
at the highest transverse momentum. There is thus a strong sensitivity to 
the symmetry energy at high transverse momenta at 30 MeV/A. The n/p ratio
differs by more than a factor of 2 by using $\gamma=0.5$ and $\gamma=2$.
This is mainly due to the fact that these high transverse momentum nucleons 
have gone through regions of higher density gradients. 

At 400 MeV/A, effects of the symmetry potential is still obvious but not as dramatic as 
that at 30 MeV/A. Two effects are responsible for this observation. First, 
mean field effects are weaker at higher energies than at lower energies. 
Secondly, because of the particular 
functional form of the symmetry potential used here, the relative strengthes 
of the symmetry potentials are just opposite at supranormal and 
subnormal densities with $\gamma=0.5$ and 2 as shown in Fig.\ 1.
At 30 MeV/A, only the low density part up to about $1.2\rho_0$ contributes 
while at 400 MeV/A a broad range of densities up to about $3\rho_0$ is involved. 
Of course, at 400 MeV/A effects of the high density region dominate although
there is a concealing effect for particles going through both low and high density 
regions. Thus the n/p ratio is higher with $\gamma=2$ than with $\gamma=0.5$ in contrast to the
30 MeV/A case. Therefore, the mid-rapidity 
n/p ratio observable is mostly useful for learning the symmetry 
energy at subnormal densities in reactions with beam energies around the Fermi energy.  
We shall discuss in the following two observables that are more useful for learning the 
symmetry energy at supranormal densities in reactions at higher beam energies.    
    
\begin{figure}[htp] 
\centering \epsfig{file=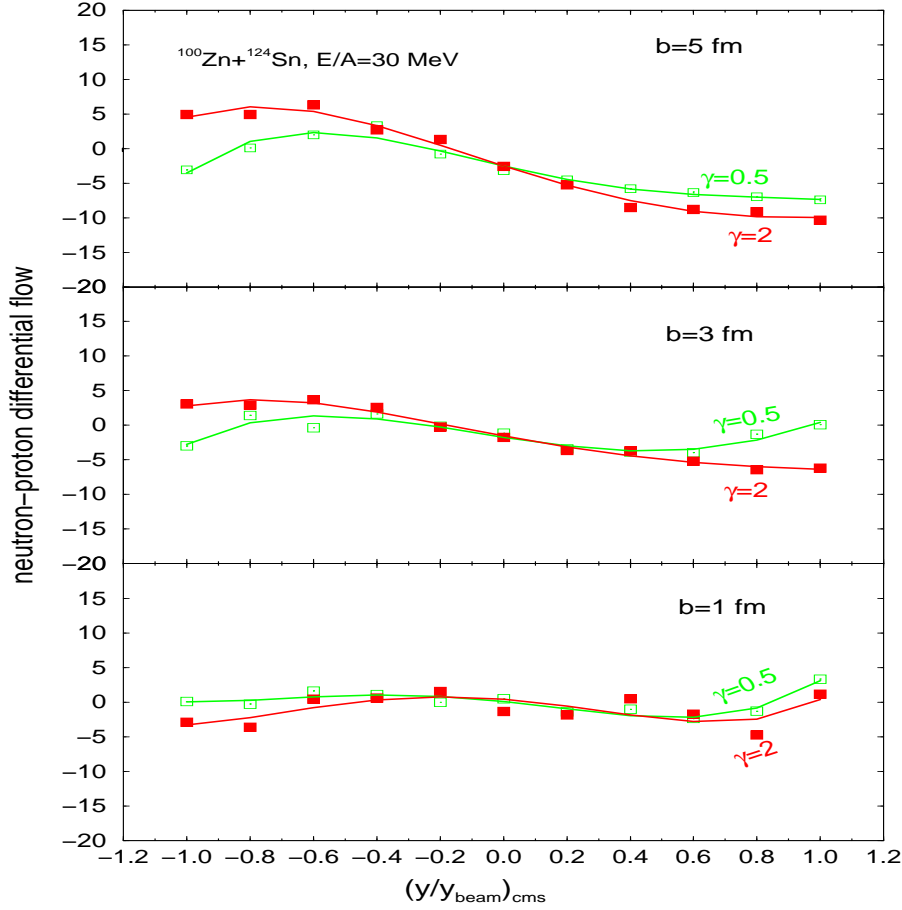,width=12cm,height=12cm,angle=-90}
\vspace{1.0 cm}  
\caption{Neutron-proton differential flow for the reaction of 
$^{100}Zn+^{124}Sn$ at a beam energy of 30 MeV/A.} 
\label{fig8} 
\end{figure}     
We now examine the neutron-proton differential flow for the neutron-rich system 
$^{100}Zn+^{124}Sn$. The neutron-proton differential flow in 
neutron-deficient reactions is not sensitive to the symmetry energy as one expects. 
The neutron-proton differential flow is defined as
\begin{equation}
F^x_{n-p}(y)\equiv\sum_{i=1}^{N(y)}(p^x_iw_i)/N(y),
\end{equation}
where $w_i=1 (-1)$ for neutrons (protons) and $N(y)$ is the total number of free 
nucleons at rapidity $y$. It combines constructively effects of the 
symmetry potential on the isospin fractionation and the collective flow. 
It also has the advantage of maximizing effects of the symmetry potential 
while minimizing effects of the isoscalar potential\cite{li00}. 
Shown in Fig. 8 and Fig. 9 are the neutron-proton differential flows 
in central and mid-central collisions at 30 MeV/A and 400 MeV/A, respectively. 

\begin{figure}[htp] 
\centering \epsfig{file=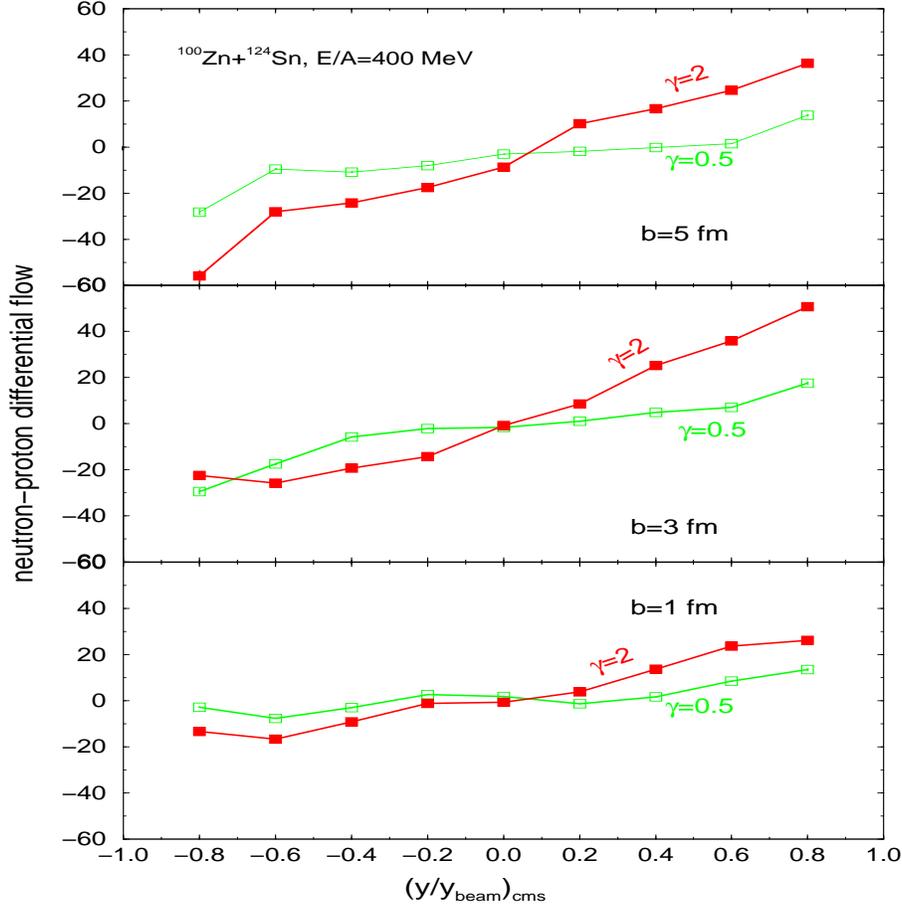,width=12cm,height=12cm,angle=-90}
\vspace{1.0 cm}  
\caption{Neutron-proton differential flow for the reaction of 
$^{100}Zn+^{124}Sn$ at a beam energy of 400 MeV/A.} 
\label{fig9} 
\end{figure}     

At 30 MeV/A, one observes an anti-flow while at 400 MeV/A there is 
a positive flow. This transition from negative to positive flow from the
Fermi energy to relativistic energies is well known\cite{gary}. At both energies 
there is clearly an observable symmetry energy effects. Again, the 
higher magnitude of symmetry potential at low densities with $\gamma=0.5$
and at higher densities with $\gamma=2$ leads to the higher neutron-proton
differential flow at 30 MeV/A and 400 MeV/A, respectively. 
Obviously the symmetry energy effect is now much stronger at 400 MeV/A than at 30 MeV/A.
This is consistent with the well-known finding that the transverse flow 
is sensitive to the early density and pressure gradients\cite{science}. 
Therefore, the neutron-proton differential flow is more useful for studying 
the symmetry energy at high densities.   
      
\begin{figure}[htp] 
\centering \epsfig{file=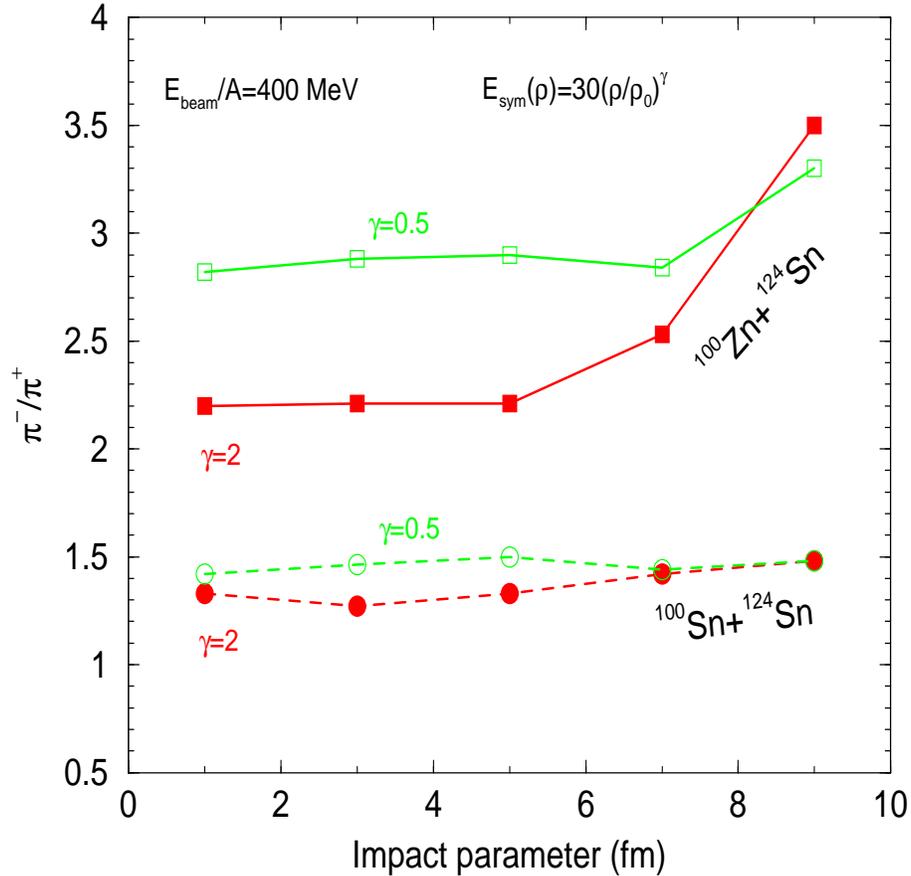,width=12cm,height=12cm,angle=-90}
\vspace{1.0 cm}  
\caption{$\pi^-/\pi^+$ ratio as a function of impact parameter in the 
reaction of $^{100}Zn+^{124}Sn$ at a beam energy of 400 MeV/A.} 
\label{fig10} 
\end{figure}
Another observable known to be sensitive to the high density behaviour of symmetry energy
is the $\pi^-/\pi^+$ ratio\cite{li02,li03}. Shown in Fig. 10 are the $\pi^-/\pi^+$ 
ratios at 400 MeV/A from central to 
peripheral collisions. It is seen that the $\pi^-/\pi^+$ ratio is sensitive to
both the n/p ratio of the reaction system and the symmetry energy, especially
in central collisions. As discussed in detail in ref.\cite{li02,li03}, 
the $\pi^-/\pi^+$ ratio is sensitive to the isospin asymmetry of the high density
region $(n/p)_{dense}$. While the latter is determined by the stiffness of the 
symmetry energy. The soft symmetry energy at supranormal densities favors the 
formation of a neutron-rich high density region and a neutron-deficit 
low density region. In our discussions above, it has been shown that at 400 MeV/A 
with $\gamma=0.5$ the n/p ratios of free nucleons are less than those with $\gamma=2$.
Thus the corresponding $(n/p)_{dense}$ and $\pi^-/\pi^+$ ratios are 
higher with $\gamma=0.5$. It is seen from Fig. 10 that the $\pi^-/\pi^+$ ratio is
about 25\% higher with $\gamma=0.5$ than with $\gamma=2$ in central collisions. 

\section{Summary}
In summary, within an isospin-dependent transport model for nuclear reactions 
induced by neutron-rich nuclei, we carried out a comparative study of $^{100}Sn+^{124}Sn$ 
and $^{100}Zn+^{124}Sn$ reactions at a beam energy of 30 MeV/A and 400 MeV/A, respectively.
The strength of isospin transport/diffusion is studied by using the rapidity 
distributions of free nucleons and their isospin asymmetries. An approximate isospin 
equilibrium can be obtained even for peripheral collisions with only the  
soft symmetry energy at 30 MeV/A. At 400 MeV/A, however, an appreciable isospin 
translucency happens even in the most central collisions. We found that 
the neutron/proton ratio at mid-rapidity but high transverse momenta 
at 30 MeV/A is most sensitive to the symmetry energy at subnormal densities. 
On the other hand, at higher beam energies, e.g., 400 MeV/A, 
neutron-proton differential flow and the $\pi^-/\pi^+$ ratio are 
found to be more sensitive to the density-dependence of the symmetry energy at 
supranormal densities. To map out the entire density-dependence of the 
symmetry energy a combination of several complementary observables in 
a broad range of beam energies is necessary. 
These observations are useful for planning experiments and 
designing new detectors for exploring the EOS of neutron-rich 
matter at RIA. Several observables useful for studying the EOS of 
neutron-rich matter require the detection of neutrons. 

We would like to thank Dr. J. Dobaczewski and Dr. Lie-Wen Chen for providing us 
the initial nucleon density distributions used in this work. 
This work was supported in part by the National Science Foundation 
Grant No. PHY-0088934 and PHY-0243571.

\end{document}